\documentclass[twocolumn,aps,prl,floatfix]{revtex4}
\usepackage[T1]{fontenc}
\usepackage[latin9]{inputenc}
\usepackage{amsmath}
\usepackage{amssymb}
\usepackage{graphicx}
\usepackage{esint}

\makeatletter

\@ifundefined{textcolor}{}
{%
 \definecolor{BLACK}{gray}{0}
 \definecolor{WHITE}{gray}{1}
 \definecolor{RED}{rgb}{1,0,0}
 \definecolor{GREEN}{rgb}{0,1,0}
 \definecolor{BLUE}{rgb}{0,0,1}
 \definecolor{CYAN}{cmyk}{1,0,0,0}
 \definecolor{MAGENTA}{cmyk}{0,1,0,0}
 \definecolor{YELLOW}{cmyk}{0,0,1,0}
}

\makeatother

\begin{document}

\title{Amplitude control of quantum interference}

\author{W. J. Mullin$^{a}$ and F. Laloë$^{b}$}

\affiliation{$^{a}$Department of Physics, University of Massachusetts, Amherst,
Massachusetts 01003 USA\\
 $^{b}$ Laboratoire Kastler Brossel, ENS, UPMC, CNRS; 24 rue Lhomond,
75005 Paris, France}

\email{mullin@physics.umass.edu;laloe@lkb.ens.fr}

\begin{abstract}
Usually, the oscillations of interference effects are controlled by
relative phases. We show that varying the amplitudes of quantum waves,
for instance by changing the reflectivity of beam splitters, can also
lead to quantum oscillations and even to Bell violations of local
realism. We first study theoretically a generalization of the Hong-Ou-Mandel
experiment to arbitrary source numbers and beam splitter transmittivity.
We then consider a Bell type experiment with two independent sources,
and find strong violations of local realism for arbitrarily large
source number $N$; for small $N$, one operator measures essentially
the relative phase of the sources and the other their intensities.
Since, experimentally, one can measure the parity of the number of
atoms in an optical lattice more easily than the number itself, we
assume that the detectors measure parity. 
\end{abstract}
\maketitle
\emph{I. Introduction.} In classical and quantum physics, the usual
control parameter of interference phenomena is the phase.\ For instance,
the interference pattern observed on a screen occurs because, at the
various points of the screen, the fields radiated from two coherent
sources have variable phase differences. In classical physics, this
is explained by the usual Fresnel construction in the complex plane,
where the phase difference controls the angle between two vectors,
leading to oscillations as a function of this phase \cite{Born-and-Wolf};
by contrast, no oscillation is expected when the amplitude of the
vectors is changed at constant phase.\ In quantum physics, the phase
also often plays the role of a parameter controlling oscillations,
e.g., at the output of a Mach-Zhender interferometer \cite{Mach}
crossed by a series of single particles.\ Another example is the
oscillations of correlation functions leading to the observation of
violations of Bell inequalities, where the control parameters are
the rotation of linear analyzers defining the relative phase of two
circular polarizations \cite{Freedman}.\ The purpose of this article
is to show that, in quantum physics, changing the amplitudes can also
lead to strong oscillations and quantum interference effects. These
oscillations occur with bosonic systems, which can be described either
as fields or systems of particles.\ Curiously, they are due to the
particle character of the quantum system, and disappear when the granularity
of the field vanishes and when detectors measure continuous intensity
variables \cite{semiclassical}. 

A motivation for this study is given by recent experiments made with
Bose-Einstein condensates and atomic interferometers with ultracold
gases \cite{AtomInterf1}.\ Atom beam splitters \cite{AtomInterf1}
may either involve Bragg scattering \cite{Bragg1,Bragg2} or be formed
by the use of radio-frequency-induced adiabatic double-well potentials
\cite{Kruger}. In the latter case, the splitting of one condensate
into two parts can easily be adjusted to provide various given ratios
between their populations, corresponding naturally to beam splitters
with variable transmission and reflection coefficients.\ Moroever,
recent experiments using optical lattices have shown that, while counting
individual particles may be difficult, one\
can much more easily measure the parity of the number of particles
trapped in a potential well \cite{Greiner,Cheneau}. The reason is
that, on each lattice site, atoms recombine by pairs and form molecules
escaping the trap. This is why we study the effect of beam splitters
with variable transmittivity on the parity of the number of particles
in each output beam. While we emphasize the use of ultracold gases in the experiments we propose,  it may be possible to produce the necessary
Fock states  by photonic methods
\cite{Hof,Wang,Geremia,Dots}. 

In this paper we discuss two possible experiments: one with two sources
and one beam splitter and two detectors, the other with more beam
splitters and detectors and illustrating quantum non-locality. The
first is a simple generalization of the Hong-Ou-Mandel (HOM) \cite{Hong-Ou-Mandel}
experiment in which two bosons (photons or atoms) interfere at a beam
splitter, resulting in the absence of any possible coincidence counts
in the two detectors. Here we consider arbitrary source populations
and the effect of changing the reflectivity of the beam splitter.\ In
the second, we extend violations of the Bell inequalities, found previously
\cite{EuroLM,FL2} with Fock-state condensates, to cases where the
reflectivities are used as control parameters; indeed we find that
the violations actually exceed those obtained by controlling phase
shifts. 
\begin{figure}[h]
\centering \includegraphics[width=2in]{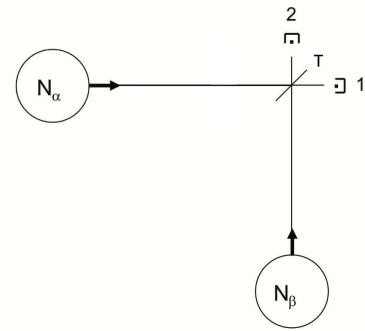} \caption{$N_{\alpha},N_{\beta}$ bosons proceed from the sources to a beam
splitter, followed by two detectors 1 and 2, where $m_{1}$ and $m_{2}$
particles are detected. The beam splitter has an adjustable transmission
coefficient, $T$, not necessarily set to $1/2$. }

\label{fig1-1} 
\end{figure}

\emph{II. Generalized HOM Effect and Parity}. We generalize the HOM
effect to an arbitrary number of photons and to arbitrary $T$ and
$R=1-T$, using the same formalism as in \cite{foundLM} (where R
and T were each taken equal to 1/2). We also study whether such a
generalized HOM experiment (GHOM) can be performed if only the measurement
of the parity of the numbers of the particles at the detectors is
available. The device is shown in Fig.~\ref{fig1-1}.

\begin{figure}[h]
$\centering$ \includegraphics[width=2.5in]{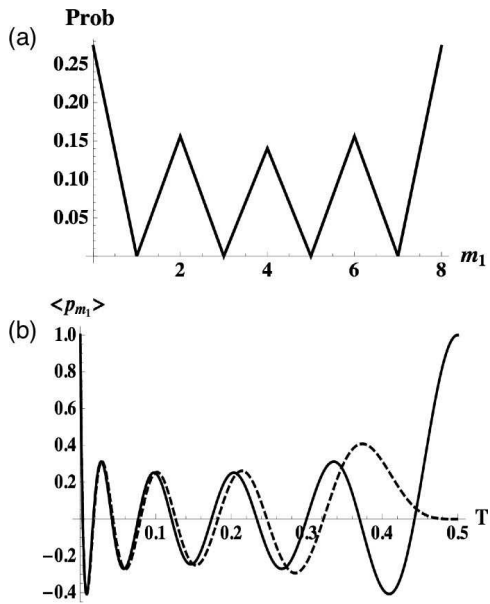} \caption{(a) The probability of Eq.~(\ref{eq:QAform}) vs. $m_{1}$ for $N_{\alpha}=4,\text{~}N_{\beta}=4$
and $T=R=1/2$ illustrating the rule that, if an even number of particles
enters each side of the beam splitter, an even number must emerge
from each side. (b) Parity average versus $T$ for $N_{\alpha}=N_{\beta}=10$
(solid) and $N_{\alpha}=12,$ $N_{\beta}=8$ (dashed). }

\label{fig:GHOM} 
\end{figure}

Before the beams of bosons cross the beam splitter, they are described
by the quantum state
\begin{equation}
\left\vert N_{\alpha},N_{\beta}\right\rangle =\frac{1}{\sqrt{N_{\alpha}!N_{\beta}!}}a_{\alpha}^{\dagger N_{\alpha}}a_{\beta}^{\dagger N_{\beta}}\left\vert \text{0}\right\rangle \label{initialstate}
\end{equation}
The destruction operators associated with the two output beams (and
detectors) are 
\begin{equation}
a_{1}=\left(\sqrt{T}a_{\alpha}+i\sqrt{R}a_{\beta}\right);\qquad a_{2}=\left(i\sqrt{R}a_{\alpha}+\sqrt{T}a_{\beta}\right)\label{def-a}
\end{equation}
The amplitude for finding $m_{1},m_{2}$ particles in the detectors
given sources with $N_{\alpha},N_{\beta}$ particles is \begin{widetext}
\begin{align}
C_{m_{1}m_{2}}(N_{\alpha},N_{\beta}) & =\frac{1}{\sqrt{m_{1}!m_{2}!N_{\alpha}!N_{\beta}!}}\left\langle 0\right\vert a_{1}^{m_{1}}a_{2}^{m_{2}}a_{\alpha}^{\dagger N_{\alpha}}a_{\beta}^{\dagger N_{\beta}}\left\vert 0\right\rangle \nonumber \\
 & =\frac{\sqrt{N_{\alpha}!N_{\beta}!}}{\sqrt{m_{1}!m_{2}!}}\sum_{p,q}\frac{m_{1}!m_{2}!\left(\sqrt{T}\right)^{p+m_{2}-q}\left(i\sqrt{R}\right)^{q+m_{1}-p}}{p!(m_{1}-p)!q!(m_{2}-q)!}\delta_{p+q,N_{\alpha}}\delta_{m_{1}+m_{2}-p-q,N_{\beta}}\nonumber \\
 & =\sqrt{\frac{N_{\alpha}!N_{\beta}!}{m_{1}!m_{2}!}}\int_{-\pi}^{\pi}\frac{d\phi}{2\pi}e^{-iN_{\alpha}\phi}\left(\sqrt{T}e^{i\phi}+i\sqrt{R}\right)^{m_{1}}\left(i\sqrt{R}e^{i\phi}+\sqrt{T}\right)^{m_{2}}\label{eq:C}
\end{align}
where we have replaced the first $\delta$-function of the second
line in Eq.~(\ref{eq:C}) by $\int\frac{d\phi}{2\pi}e^{i\phi(p+q-N_{\alpha})}$,
and have redone the sum. The square of the modulus of this expression
contains an integral over two variables $\phi$ and $\phi^{\prime}$;\ if
we make the changes of variables $\lambda=(\phi+\phi^{\prime}+\pi)/2$
; $\Lambda=(\phi-\phi^{\prime})/2$, we find for the probability the
expression:
\begin{eqnarray}
P(m_{1},m_{2}) & = & \frac{N_{\alpha}!N_{\beta}!}{m_{1}!m_{2}!}\int_{-\pi}^{\pi}\frac{d\lambda}{2\pi}\int_{-\pi}^{\pi}\frac{d\Lambda}{2\pi}e^{-i(N_{\alpha}-N_{\beta})\Lambda}\left[Te^{i\Lambda}+Re^{-i\Lambda}-2\sqrt{TR}\cos\lambda\right]^{m_{1}}\\
 &  & \times\left[Re^{i\Lambda}+Te^{-i\Lambda}+2\sqrt{TR}\cos\lambda\right]^{m_{2}}\label{eq:QAform}
\end{eqnarray}

That this probability shows interference effects is seen in Fig. \ref{fig:GHOM}(a)
for the case of $T=R=1/2$ and $N_{\alpha}=N_{\beta}$. Only pairs
of particles reach either detector. If we define the parity as $\left\langle p_{m_{1}}\right\rangle =\sum_{m_{1}}(-1)^{m_{1}}P(m_{1},N-m_{1})$
we find unity for the case shown in Fig. \ref{fig:GHOM}(a), a first
indication that \emph{parity is a useful indicator of interference
effects}. For general values of $T$ and $R=1-T$ we find 
\begin{align}
\left\langle p_{m_{1}}\right\rangle  & =\frac{2^{N}N_{\alpha}!N_{\beta}!}{N!}\int_{-\pi}^{\pi}\frac{d\lambda}{2\pi}\int_{-\pi}^{\pi}\frac{d\Lambda}{2\pi}e^{-i(N_{\alpha}-N_{\beta})\Lambda}\left[i\left(R-T\right)\sin\Lambda+2\sqrt{TR}\cos\lambda\right]^{N}\nonumber \\
 & =2^{N}N_{\alpha}!N_{\beta}!\sum_{p=0}^{N}\frac{(-1)^{(p-N_{\alpha}+N_{\beta})/2}(R-T)^{p}\left(\sqrt{TR}\right)^{N-p}y(N-p)y(N_{\alpha}-N_{\beta}+p)}{2^{p}\left(\frac{N-p}{2}\right)!^{2}\left(\frac{N_{\alpha}-N_{\beta}+p}{2}\right)!\left(\frac{p-N_{a}+N_{\beta}}{2}\right)!}\label{eq:Gen<p>}
\end{align}
 \end{widetext}where $y(x)=\frac{1}{2}(1+(-1)^{x})$. The second
line of (\ref{eq:Gen<p>}) comes from expanding the integrand in Eq.~(\ref{eq:QAform})
and integrating term by term. For the case $T=R=1/2$ this reduces
to $\left\langle p_{m_{1}}\right\rangle _{N_{\alpha},N_{\beta}}=\delta_{N_{\alpha},N_{\beta}}$.
(While the plots of $P(m_{1},m_{2})$ for $N_{\alpha}\ne N_{\beta}$
continue to show interference effects, the probability for finding
even values of $m_{1}$ is the same as that for finding odd values,
so that parity does not show the interference in that case.) 

In Fig.~\ref{fig:GHOM}(b) we show $\left\langle p_{m_{1}}\right\rangle $
versus $T$ for equal and unequal $N_{\alpha}$ and $N_{\beta}$.
Note the values are much the same except near $T=0.5.$ To get an
understanding of the oscillations of the parity with $T$ and how
they reveal the interference effects, consider the simpler situations
where $N_{\alpha}$ and $N_{\beta}$ are small. For $N_{\alpha}=2$,
$N_{\beta}=1$ we have 
\begin{align}
a_{\alpha}^{\dagger2}a_{\beta}^{\dagger} & =\frac{1}{2^{3/2}}\left[T\sqrt{R}a_{1}^{\dagger3}+\sqrt{T}(T-2R)a_{1}^{\dagger2}a_{2}^{\dagger}\right.\nonumber \\
 & \left.+\sqrt{R}(R-2T)a_{1}^{\dagger}a_{2}^{\dagger2}-Ra_{2}^{\dagger2}\right]
\end{align}
 From this we see that negative parity is favored when $T=2R$ , ($T=0.66)$
and positive for $R=2T$, ($T=0.33)$ and this is very close to what
we find by explicit calculation. Again we have cancellation for the
various possible ways two particles can get to detector 1 and one
to detector 2 and vice versa. These maxima and minima estimates are
not exact since the parity depends on all processes, not just a subset.
Sanaka et al \cite{Sanaka} have considered the special case where
$N_{\alpha}=n$ and $N_{\beta}=1$ and shown that $P(1,n)$ of Eq.~(4)
vanishes when $R=n/(n+1)$ allowing filtering of $n$-particle states
out of an input beam. If parity is more easily measured than actual
detector counts, one could argue that the same is true of source numbers.
For the case $T=1/2$ a random distribution of source numbers $N_{\alpha},N_{\beta}$
will favor even parity because of the occasional occurrence of terms
where $N_{\alpha}=N_{\beta}$. With a binomial source distribution,
where the total number of particles is known to be $N$, we have a
source-averaged parity of 
\begin{eqnarray}
\left\langle p_{m_{1}}\right\rangle  & = & \sum_{N_{\alpha=0}}^{N}\frac{N!\delta_{N_{\alpha},N-N_{\alpha}}}{{2}^{N}N_{\alpha}!(N-N_{\alpha})!}=\begin{cases}
\frac{N!}{2^{N}(N/2)!^{2}}\\
0
\end{cases}
\end{eqnarray}
 where the top line holds for $N$ even and the bottom for $N$ odd.
(The latter result holds because we cannot have $N_{\alpha}=N_{\beta}$
with odd total $N$). An analogous result will hold for any source
distribution. If we can count the parity of the total source distribution,
we can always guarantee to see the interference result. As $N$ increases
the average parity decreases, but the method works well for small
$N.$ Analogous arguments can be made for $T\neq1/2.$

Parity oscillations, as a function of $T$, therefore provide a useful
signature of the GHOM quantum effect. We now show that the same ideas
can lead to strong violations of the Bell-Clauser-Horne-Shimony-Holt
(BCHSH) inequalities \cite{Bell,CHSH}.

\begin{figure}[h]
\centering \includegraphics[width=3in]{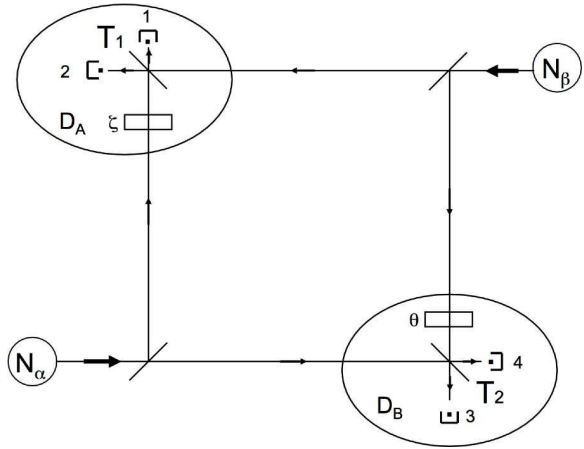} \caption{Two Fock states, with populations $N_{\alpha}$ and $N_{\beta}$,
divide at beam splitters, and are made to interfere in two regions
$D_{A}$ and $D_{B}$, with counting in detectors 1 and 2 in the former,
3 and 4 in the latter.\ The phase shifts, $\zeta$ and $\theta$,
are zero in this analysis. The transmission coefficients at $D_{A}$
and $D_{B}$, $T_{1}$ and $T_{2}$, respectively, are varied. The
transmission coefficients at the sources remain set at 1/2. }

\label{fig:Bell Interf} 
\end{figure}

\emph{III. Violating BCHSH inequalities by varying transmission coefficients.}
The interferometer we analyze is shown in Fig.~\ref{fig:Bell Interf}.
We have analyzed this device previously \cite{FL2,EuroLM} with variations
of the phase shifters and have seen that the Bell inequalities may
be violated for arbitrarially large $N$. In the present analysis
we want to allow the experimenters, Alice and Bob, to vary the transmission
coefficients $T_{1}$ and $T_{2}$ at their detectors.

The corresponding operators are 
\begin{equation}
\begin{array}{l}
a_{1}=\frac{i}{\sqrt{2}}\left[\sqrt{T_{1}}a_{\alpha}+\sqrt{R_{1}}a_{\beta}\right]\text{;\ }a_{2}=\frac{-1}{\sqrt{2}}\left[\sqrt{R_{1}}a_{\alpha}-\sqrt{T_{1}}a_{\beta}\right]\\
a_{3}=\frac{i}{\sqrt{2}}\left[\sqrt{R_{2}}a_{\alpha}+\sqrt{T_{2}}a_{\beta}\right]\text{;\ }a_{4}=\frac{1}{\sqrt{2}}\left[\sqrt{T_{2}}a_{\alpha}-\sqrt{R_{2}}a_{\beta}\right]
\end{array}
\end{equation}
 or generally $a_{i}=u_{i}a_{\alpha}+v_{i}a_{\beta}$. We consider
the case where the sources are equal: $N_{\alpha}=N_{\beta}=N/2$.
By proceeding as we did above we find the probability for finding
$\{m_{1},m_{2},m_{3},m\}$ is 
\begin{align}
P_{m_{1}m_{2}m_{3}m_{4}} & =\frac{(N/2)!^{2}}{m_{1}!\cdots m_{4}!}\int\frac{d\phi^{\prime}}{2\pi}e^{iN\phi^{\prime}/2}\nonumber \\
 & \times\int_{-\pi}^{\pi}\frac{d\phi}{2\pi}e^{-iN\phi/2}\prod_{i=1}^{4}\Omega_{i}^{m_{i}}
\end{align}
 where
\begin{equation}
\Omega_{i}=\left(u_{i}^{*}e^{-i\phi^{\prime}}+v_{i}^{*}\right)\left(u_{i}e^{i\phi}+v_{i}\right)
\end{equation}

\begin{figure}[h]
\centering \includegraphics[width=3in]{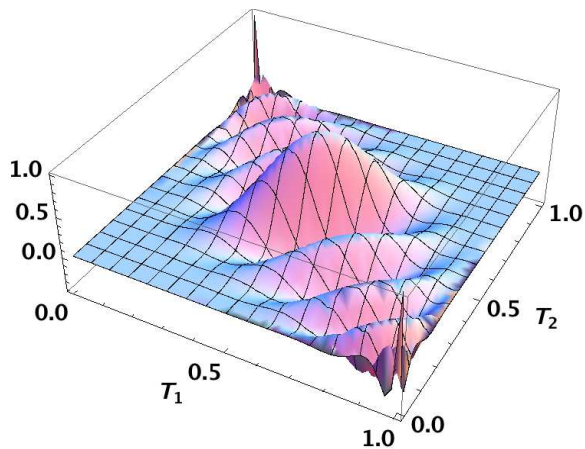} \caption{(Color online) Oscillations in $\left\langle \mathcal{AB}\right\rangle $
as a function of $T_{1}$ and $T_{2}$. $N_{\alpha}=N_{\beta}=10$.}

\label{fig:ABwiggles} 
\end{figure}

For the parity correlation we want the average of $\mathcal{AB}$
where $\mathcal{A}=(-1)^{m_{2}}$ and $\mathcal{B}=(-1)^{m_{4}}$.
After a straightforward calculation we find 
\begin{equation}
\left\langle \mathcal{AB}\right\rangle =\left(\frac{N}{2}!\right)^{2}\sum_{p=0,2,}^{N}\frac{(-1)^{p/2}\Delta T^{p}\tau^{N-p}}{(\frac{p}{2})!^{2}(\frac{N-p}{2})!^{2}}\label{eq:ABsum}
\end{equation}
 where 
\begin{eqnarray}
\Delta T & = & T_{1}-T_{2}\\
\tau & = & \sqrt{T_{1}(1-T_{1})}+\sqrt{T_{2}(1-T_{2})}
\end{eqnarray}
 If we plot $\left\langle \mathcal{AB}\right\rangle $ as a function
of $T_{1}$ and $T_{2}$ (Fig.~\ref{fig:ABwiggles}) we find oscillations
analogous to to those in Fig.~\ref{fig:GHOM}. 

\begin{figure}[h]
\includegraphics[width=3in]{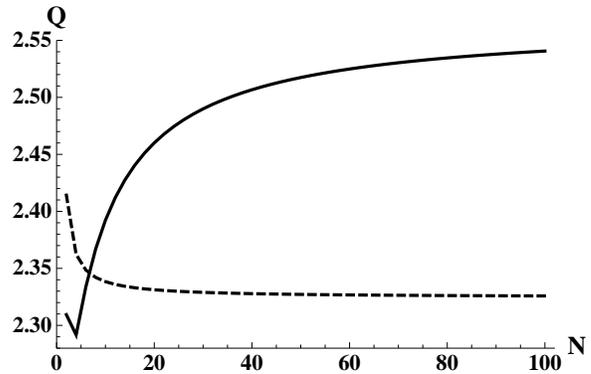}
\caption{ BCHSH Quantity $Q$ vs. $N$. The solid line is the result
of varying the transmission coefficients. The dashed line compares
the resuts from Refs. \cite{EuroLM,FL2} in which the phase shifts
were varied. }
\label{TBell}
\end{figure}

The BCHSH inequality \cite{CHSH} is 
\begin{equation}
Q=\left\langle \mathcal{AB}\right\rangle +\left\langle \mathcal{AB}^{\prime}\right\rangle +\left\langle \mathcal{A}^{\prime}\mathcal{B}\right\rangle -\left\langle \mathcal{A}^{\prime}\mathcal{B}^{\prime}\right\rangle \leq2
\end{equation}
 where the primes refer to using the four pairs of variables $T_{1}$,
$T_{2},$ $T_{1}^{\prime}$, and $T_{2}^{\prime}$. For $N=2$ we
find a maximum of $Q=2.31$ for the set of $T$ values $\{0.57,$
0.43, 0.06, 0.94\}. As $N$ increases the optimal $Q$ increases and
$T$ values move close to $1/2$. A plot is shown in Fig.~\ref{TBell}
\begin{figure}[h]
\includegraphics[width=3in]{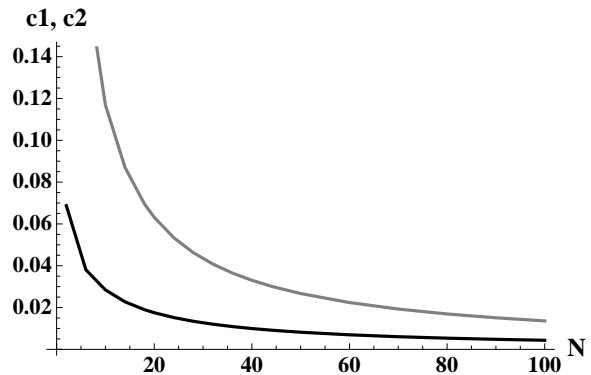}
\caption{The maximization parameters $c_{1}$ (in black)
and $c_{2}$ (in gray) giving the differences of the transmission
coefficients from 0.5 at maximum. }
\label{CBell}
\end{figure}
The $Q$ value found at $N=100$ is $2.54$ with the $T$ set of \{0.486,
0.504, 0.514, 0.486\}. The maximum possible value extrapolates to
$\sim2.56$ at large $N.$ The $T$ values range around 0.5 in terms
of just two variables $c_{1}$ and $c_{2}$, as follows $\{T_{1},T_{2},T_{1}^{\prime},T_{2}^{\prime}\}=\{0.5-c_{1},0.5+c_{1},0.5+c_{2},0.5-c_{2}\}$.
(See Fig. \ref{CBell}.)  For very small $N$ we have $T_{1}^{\prime}$, and $T_{2}^{\prime}$
near 0 and 1, respectively. The four detectors register $m_{1}\cdots m_{4}$
from which, in a second step, one calculates two parities. If $T_{1}$
and $T_{2}$ are $1/2$, no detector can distinguish the source from
which the particles originate; the ratios between the $m_{i}$ provide,
classically, the relative phase of the sources. If the $T$ values
are 0 or 1, the source populations are directly measured. Thus for
very small $N$ our scheme involves a combination of experiments where
Alice and Bob essentially measure, either the relative phase (with
$T_{1}$ and $T_{2}$ near $1/2$), or the source numbers (with $T_{1}$
and $T_{2}$ near 0 or 1). The conjugate variables here are numbers
and phase, instead of the usual quadrature operators in Bell violations.

It is interesting to compare (see Fig.~\ref{TBell}) our results  to the case in Refs.~\cite{EuroLM,FL2}
where we varied the phase shifts ($\zeta$ and $\theta$ in Fig.~\ref{fig:Bell Interf}).
 There we had $Q=2.41$ at $N=2$ with $Q$
then decreasing until it reached a limit of $2.32$ at large N. With
phase-angle variation, $Q$ decreases with $N$, but with $T$ variation
it increases with $N$ and becomes much larger than occurred with
the angle variation.

We tried varying both angles and transmission coefficients simultaneously
using four pairs of variables $\{T_{1},\zeta\}$, $\{T_{2},\theta\},$
$\{T_{1}^{\prime},\zeta^{\prime}\}$, and $\{T_{2}^{\prime},\theta^{\prime}\}$.
We never succeeded in improving the results.

Parity, which is a possible measurable variable in ultra-cold gases,
provides a useful signature of quantum interference and non-local
effects. The more surprising result of our analysis is that, even
for systems with a large number of particles, the probability of particle
transmission provides a powerful way of observing these phenomena.
In studying the GHOM effect we find curious oscillations of the parity
as $T$ is varied. In Bell violations the wave amplitude variation
actually achieves greater violations than by changes in phase.

\ Laboratoire Kastler Brossel is ``UMR 8552 du CNRS, de l'ENS, et
de l'Université Pierre et Marie Curie\textquotedblright{}.

\end{document}